\documentstyle[preprint,aps,12pt]{revtex}
\begin{document}
\draft
%\preprint{This is compuscript ``multinpi''(=BODU1)}
\date{\today}
\preprint{\Large E2 - 95 - 503}
%Reference Number:
\title{MULTI-INSTANTON EFFECTS IN QCD SUM RULES FOR THE
       PION\footnote{Work supported in part by the Heisenberg-Landau
                     program} \\}
\author{A.\ E.\ Dorokhov,
        S.\ V.\ Esaibegyan\footnote{On leave of absence from
        Yerevan Physics Institute, Yerevan, Armenia},
        N.\ I.\ Kochelev,
        and
        N.\ G.\ Stefanis\footnote{On leave of absence from
        Institut f\"ur Theoretische Physik~II,
        Ruhr-Universit\"at Bochum, \\
        \phantom{+} D-44780 Bochum, Germany}}
\address{Bogoliubov Laboratory of Theoretical Physics,
         Joint Institute for Nuclear Research, \\
         141980 Dubna, Moscow Region, Russia \\}
\maketitle
%\newpage
\begin{abstract}
Multi-instanton contributions to QCD sum rules for the pion are
investigated within a framework which models the QCD vacuum as an
instanton liquid. It is shown that in singular gauge the sum of
planar diagrams in leading order of the $1/N_{c}$ expansion
provides similar results as the effective single-instanton
contribution. These effects are also analysed in regular gauge.
Our findings confirm that at large distances the correlator
functions are more adequately described in the singular gauge
rather than in the regular one.
\end{abstract}
\newpage  %finishes title-page

The QCD sum rule approach~\cite{SVZ79} allows the investigation of
hadron properties in a systematic manner. It provides, in particular,
the possibility to describe static characteristics of particles, such
as masses, decay constants, form factors, etc., in an energy region
where perturbative methods are not
applicable~\cite{Iof81,BI82,CDKS82,IS83,NR82}.
Within this approach, the effects of large distances are effectively
parametrized in terms of local matrix elements of quark-gluon
operators averaged over the physical vacuum (vacuum condensates),
quantities which are independent of hadron properties.
On the other hand, short-distance physics is contained in the Wilson
coefficients of the Operator Product Expansion (OPE) entering the
calculation of correlators.

However, the nonperturbative matrix elements contain contributions
that are not taken into account in the OPE. These are, so-called,
``direct'' small-size instantons (see, e.g.,~\cite{NSVZ82}) which
give essential nonlocal contributions to current correlators in the
channels where they are allowed by quantum numbers. These
contributions are not sufficiently accounted for in the {\it local}
condensates, since the latter correspond to vacuum fluctuations with
infinite correlation length. They should rather be taken into account
within the Wilson coefficients along with the factors calculated by
perturbative methods.

The instanton liquid model of the QCD vacuum, originally suggested
in~\cite{Shu82,Shu83}, has later been further generalized by an
analytic approach, based on the Feynman variational
principle~\cite{DP84,DP86,DP85}. (For lattice calculations
using this vacuum model, see, e.g.,~\cite{CH92}.)

As it was shown in~\cite{DP85}, the instanton-induced vacuum
fluctuations are responsible for the spontaneous breaking
of chiral invariance. This chiral-symmetry-breaking mechanism is
based on the idea of mixing and delocalization of fermion zero modes
in the field of the instanton ($I$) and anti-instanton ($A$) pairs.
The QCD vacuum is modeled as an $I-A$ diluted liquid, characterized
by a small ratio
$\rho _{c}/R \simeq 1/3$,
where
$\rho _{c}\simeq 1/600$~MeV$\simeq 1/3$~Fm
is the average instanton size in the vacuum, and $R$ is the average
distance between pseudoparticles.

A summary of some successful applications of this approach includes:
The calculation of current correlators in the background of $I$ and
$A$ external fields which provides a useful procedure for extracting
the static features of the pseudoscalar meson
octet~\cite{Shu82,ET89}.
More recently~\cite{ET95}, a possible mechanism for the bound-state
formation in the vector-meson channel has been proposed.
In a series of works~\cite{Koc85}, several main properties of hadron
spectroscopy have been quantitatively determined.
Evidence was provided there that large spin-flip high-energy
amplitudes~\cite{DKZ93} are the result of the spin-dependent
interaction between quarks, induced by the small-size vacuum
fluctuations.

The role of direct instantons in stabilizing the QCD sum rules for
the nucleon~\cite{Iof81,BI82,CDKS82} was first discussed
in~\cite{DK90} and later also in~\cite{FB93}. These analyses show
that the inclusion of the instanton contributions amount to a
significant enlargement of the stability region of the Borel
parameter.

The instanton contribution to different vacuum matrix elements
is defined basically by the quark zero modes in external $I\;,\;A$
fields. Due to the specific chiral and flavor properties of these
fields, instanton effects depend strongly on the channel under
consideration. In the channel with the quantum number $0^{-}$, the
instanton contribution is dominant~\cite{NSVZ82}. The single
instanton contribution to the QCD sum rule for the pion, within the
effective approach given in~\cite{Shu83,SVZ80}, has been first
calculated in~\cite{Shu83}. He has shown that a self-consistent
description of the pion as a pseudo-Goldstone mode is possible only
if the contribution of direct instantons is taken into account.

It is the purpose of this paper to investigate the multi-instanton
contributions to QCD sum rules for the pion in the framework proposed
in~\cite{DP85}.
The main conclusion of this investigation is that the large-distance
behavior of the pion correlator in the singular gauge is essentially
the same as in the effective single instanton approach~\cite{Shu83}.
The behavior of the correlator in the regular gauge is also
explored but found to give a negligible contribution at large
distances.

The QCD sum rules for the pion are evaluated from the correlator
function
\begin{equation}
  \Pi (q)
=
  i \int d^{4}x \,{\rm e}^{iqx}
    \langle 0|T(j(x)j^{+}(0)|0 \rangle \; ,
\label{E1}
\end{equation}
%Eq (1) Pion correlator
which is considered at $ Q^{2}=-q^{2}\simeq 1$~GeV.
We will analyze the sum rules for a charged pion, so that
\begin{equation}
  j(x)
=
   q_{u}q_{\bar d}\left[ \bar v_{R}i\gamma _{5}u_{L}
 + \bar v_{L}i\gamma _{5}u_{R}\right](x) \; .
\label{E2}
\end{equation}
%Eq (2) Charged pion current
Here $q_i$ denotes quark annihilation operators, and
$
 u_{L(R)}
=
 \left ( \frac{1\pm\gamma _5}{2} \right)
$
are left- (right-) handed spinors.

The single instanton contribution has been computed in~\cite{Shu83},
assuming that the quark Green function in the background of the
instanton field
\begin{equation}
  S_{I}(x,y)
=
  S_{0}(x,y)+S_{\pm}(x,y)
\label{E3}
\end{equation}
%Eq (3) Quark Green function in the instanton field
can be approximated by the expression
\begin{equation}
  S_{\pm}(x,y)
=
  \langle
         q^{a}_{\alpha}(x) {\bar q}^{b}_{\beta}(y)
  \rangle
=
  \int d^{4}z \,
  \frac{\left[
            \Psi ^{\pm}_{z}(x){\bar\Psi}^{\pm}_{z}(y)
      \right]^{ab}_{\alpha\beta}}{m^{*}}
\label{E4}
\end{equation}
%Eq (4) Approximated quark Green function in the instanton field
which retains only the zero modes, given in singular gauge by
\begin{equation}
  \Psi ^{\pm}_{x_0}(x)
=
  \Phi (x-x_0)\frac{1\pm\gamma _5}{2} (\not\! x -\not\! x_0) U
\label{E5}
\end{equation}
%Eq (5) Zero modes
with
$
 \Phi (x)
 =
 \frac{\rho _{c}}{\sqrt{x^{2}}\pi
 \left[ x^{2} + \rho _{c}^{2} \right]^{3/2}}
$,
and where we have averaged over instanton positions, denoted by $x_0$.
In Eq.~(\ref{E4}), $(a,b)$ are color and $(\alpha ,\beta )$ spinor
indices, respectively; $U$ is the color-spin matrix $(U^{+}U=1)$,
whereas $\pm$ refers to the instanton (anti-instanton).
The effective quark mass $m^{*}$, acquired in the instanton
vacuum~\cite{Shu83,SVZ80}, is
\begin{equation}
  m^{*}
\approx
- \frac{2}{3} \pi ^{2} \langle 0|{\bar q} q| 0\rangle
  \rho _{c}^{2}\simeq 200~{\rm MeV} \; .
\label{E6}
\end{equation}
%Eq (6) Effective quark mass
Note that the free quark propagator in (\ref{E3})
\begin{equation}
  S_{0}(x,y)
=
  \frac{i({\not\! x}-{\not\! y})}{2\pi ^{2} (x-y)^{4}}
\label{E7}
\end{equation}
%Eq (7) Free quark propagator approximating non-zero modes
serves to approximately account for the contribution of the
non-zero modes.

Applying now the Borel transformation~\cite{SVZ79,Shu83}
\begin {equation}
  B[f(s)]
\equiv
  \lim_{{n\to\infty\atop s\to\infty} {n/s=\tau^{2}}}
  (-1)^{n}\,\frac{s^{n+1}}{n!}
  \left(\frac{d}{ds}\right)^{n} f(s) \; ,
\label{E8}
\end{equation}
%Eq (8) Borel transformation
and using the expression for the instanton density~\cite{Shu83,DP86},
$n_{c}(\rho )$,
\begin{eqnarray}
  n(\rho)
& = &
  n_{c}\delta (\rho -\rho _{c})
\nonumber\\
  n_{c}
& {\simeq} &
  0.8\times 10^{-3}~{\rm GeV}^{4} \; ,
\label{E9}
\end{eqnarray}
%Eq (9) Instanton properties in the instanton liquid model
derived in the instanton liquid model, in conjunction with the
relation
$
 \langle {\bar q}q\rangle
=
 - \frac{2n_{c}}{m^{*}}
$
between the instanton density and the quark condensate, we obtain
the following correlator in terms of the inverse Borel
parameter $\tau ^{2}= 1/M^{2}$
\begin{equation}
  \Pi (\tau )
=
  \frac{2n_{c}\,\rho _{c}^{2}\,\zeta}{m^{*2}\,\tau ^{4}\sqrt{\pi}}
  \int_{0}^{\infty}d\alpha\int_{0}^{\infty}d\beta \,
  {\rm e}^{-\zeta ^{2}{t^2}}
  \left(
        \zeta ^{2}t^{3}-\frac{3}{2}t
  \right)\;
        \cosh\alpha\;\cosh\beta \; ,
\label{E10}
\end{equation}
%Eq (10) Borelized correlator
where
$
 t
=
 \frac{\cosh\alpha + \cosh\beta}{2}
$, and
$
 \zeta = \rho _{c} \tau
$.
This result differs from the one given in~\cite{Shu83}
by a factor
$\sqrt{\frac{\pi}{2}}$
(which may be a misprint there).
Employing the substitutions
$
 \frac{\alpha + \beta}{2}
=
 y_{1}
$,
$
 \frac{\alpha - \beta}{2}
=
 y_{2}
$
in Eq.~(\ref{E10}), the double integral can be further expressed
via the MacDonald functions
$K_{\nu}(\frac{\zeta ^2}{2})$
to read
\begin{equation}
  \Pi (\tau )
=
  \frac{3}{8}\frac{\zeta ^{2}}{\tau ^{4}\,\pi ^{2}}\,
  {\rm e}^{-\zeta ^{2}/2}
  \left[
        K_{0}\left( \frac{\zeta ^2}{2}\right)
      +
        K_{1}\left(\frac{\zeta ^2}{2}\right)
  \right] \; ,
\label{E11}
\end{equation}
%Eq (11) Correlator via MacDonald functions

It is worth remarking again that, as it was shown in~\cite{Shu83},
the QCD sum rule in the pseudoscalar channel can be saturated only
by including the one-instanton contribution (cf.~Eq.~(\ref{E10})).

As it was in~\cite{DEK96}, the correction to $\Pi (\tau )$, arising
from the next to leading $I-A$ contribution is at the level of $5\%$
relative to the leading contribution given by Eq.~(\ref{E11}) when
$\tau$ is large, i.e.,
$\tau\simeq\rho _{c}$.
This $\tau$-region of the correlator corresponds to large distances
and hence to exploit the QCD sum rules, one may use the techniques
developed in~\cite{DP84,DP86,DP85}, based as already said on the
summation of planar diagrams in leading $1/N_{c}$ approximation.
Suffice to say that such calculations are based on a model Green
function for the quark in the background field of one instanton
which actually resembles Eq.~(\ref{E4}) with the effective quark
mass $m^{*}$ now being replaced by the current quark mass $m$.
The summation based on the delocalization mechanism of zero
modes~\cite{DP86,DP85} gives an expression for the two-point function
$\Pi (q)$ (cf. Eq.~(\ref{E1})). At small momentum $q$, the connected
part of this function defines the pion mass and the value of the
decay constant $f_{\pi}$, in good agreement with the data (see
also~\cite{ET89,Shu93,Hut95a}).

In the pseudoscalar channel, the expression for $\Pi (q)$ which
incorporates the multi-instanton/anti-instanton effects is
\begin{equation}
  \Pi (q)
=
  \frac{4\,V\,N_{c}^{2}}{N}\;
  \Gamma ^{2}_{5}(q)\;
  \frac{1}{R_{-}(q)}
\label{E12}
\end{equation}
%Eq (12) Multi-instanton effects on the correlator with $R_{-}(q)$
indicating the pole position in the $\gamma _{5}$-channel.
Here, N is the number of instantons, $ N_{c} $ the number of colors,
V the space-time volume, and $\Gamma _{5}^{2}(q)$ an effective vertex
function.
In the analysis of~\cite{DP85}, the behavior of $\Pi (q)$, expressed
in Eq.~(\ref{E12}), has been investigated in the limit ${q\to 0}$.
In the present work, we are primarily interested in the limit
$q^{2} \simeq 1$~GeV${}^{2}$, where the QCD sum rules can be safely
evaluated~\cite{SVZ79,Iof81,BI82,CDKS82,IS83}. However, in this
region of momentum the estimates derived from Eq.~(\ref{E4}) with
the replacement $m^{*} \to m$ are valid at a qualitatively level
only~\cite{DP85}. Recall that the model Green function ``works''
well in the two limiting cases
$\rho q \gg 1$ and $\rho q \ll 1$.

Then in leading ${\rho _{c}}M(0)$ approximation we have
\begin{eqnarray}
  R_{-}(q)
& = &
  \frac{2\,V\,N_{c}}{N}
  \int_{}^{}\frac{d^{4}k}{(2\pi )^{4}}\,
  \frac{M_{1}^{2}k^{2}_{2}+M_{2}^{2}k_{1}^{2}-2k_{1}k_{2}M_{1}M_{2}}
  {(M^{2}_{1}+k_{1}^{2})(M^{2}_{2}+k_{2}^{2})}
\nonumber \\
& = \atop Q^{2} \simeq 1~{\rm Gev}^{2}  &
  \frac{4\,V\,N_{c}}{N}
  \int_{}^{} \frac{d^{4}k}{(2\pi )^{4}}\,
  \frac{M^{2}(k)}{k^{2}+M^{2}(k)}
\nonumber \\
& \simeq &
  \frac{4\,V\,N_{c}~M^{2}(0)}{N}
  \int_{}^{} \frac{d^{4}k}{(2\pi )^{4}}\,
  \frac{1}{k^{2}+M^{2}(0)}
\nonumber \\
& \simeq &
  \frac{4\,V\,N_{c}\,M^{2}(0)\pi ^{2}}{(2\pi )^{4}\,N\,\rho _{c}^{2}}
\label{E13}
\end{eqnarray}
%Eq (13) Pole structure of the pion corrrelator
with
\begin{eqnarray}
  \Gamma _{5}(q)
& = &
  - \int_{}^{} \frac{d^{4}k}{(2\pi )^{4}}\,
    \frac{\sqrt{M(k)M(k+q)}\,(k^{2}+(k+q)^{2}-q^{2})}
    {[M^{2}(k)+k^{2}][M^{2}(k+q)+(k+q)^{2}]}
\nonumber\\
& {\simeq \atop Q^{2} \simeq 1~{\rm GeV}^{2}} &
 - \;\frac{\sqrt{M(0)M(Q)}
     \pi ^{2}}{(2\pi )^{4}\,2Q^{2}\,\rho _{c}^{4}} \; ,
\label{E14}
\end{eqnarray}
%Eq (14) Effective vertex function
where $M_{1(2)}$ is a short-hand notation for
$
 M(k\mp\frac{q}{2})
$.
Assuming fixed values of the instanton radii,
$
 \rho =\rho _{c}
$,
it follows
$
 M(p)\sim p^{2}\varphi ^{2}(p)
$,
$
 \varphi (p)
$,
being associated with the zero mode representation in momentum space.
It has the following asymptotics
\begin{equation}
  \varphi _{s}(p)
=
  \left\{\begin{array}{ll}
  -\frac{2\pi\rho}{|p|},
   \;\;\;\;\;\;\;\;\;\;
& \rho p\ll 1 \\
  -\frac{12\pi}{p^{4}\rho ^2},
   \;\;\;\;\;\;\;\;\;\;
& \rho p\gg 1
\end{array}
\right.
\label{E15}
\end{equation}

%Eq (15) Asymptotics of \phi_s
\begin{equation}
  \varphi_{r}(p)
=
  \left\{\begin{array}{ll}
  \frac{4\pi\rho}{|p|},
  \;\;\;\;\;\;\;\;\;\;
& \rho p\ll 1 \\
  -\frac{4\pi\rho}{p}\, {\rm e}^{-p\rho},
   \;\;\;\;\;\;\;\;\;\;
& \rho p\gg 1
\end{array}
\right.
\label{E16}
\end{equation}
%Eq (16) Asymptotics of \phi_r
in singular and regular gauges, respectively.
Since $M(p)$ is a rapidly increasing function with $p$ for
$ \rho p\gg 1 $~\cite{DP86,Hut95b}, we cut off the $k$-integration
in (\ref{E13}) and (\ref{E14}) at values $\sim$~$1/\rho ^{2}$.

Then by using Eq.~(\ref{E4}) in conjunction with
Eqs.~(\ref{E12})-(\ref{E14}), we obtain
\begin{equation}
  \Pi (q)
=
  \frac{N_{c}}{Q^{4}}\;
  \frac{M(Q)}{M(0)}\;
  \frac{1}{2^{6}\rho ^{6}\pi ^{2}} \; ,
\label{E17}
\end{equation}
%Eq (17) Cut-off expression for the correlator
and utilizing the explicit expressions for $\varphi (p)$, given
in~\cite{DP86,Hut95a}, viz.
\begin{eqnarray}
  \varphi _{s}(p)
& = &
  \pi \,\rho ^{2}\frac{d}{d\xi}
  \left[
        I_{0}(\xi )K_{0}(\xi )-I_{1}(\xi )K_{1}(\xi )
  \right]_{\xi =\frac{{\rho}p}{2}} \; ,
\nonumber \\
  \varphi _{r}(p)
& = &
  \frac{4\pi\rho}{p}\;{\rm e}^{-p\rho} \; ,
\label{E18}
\end{eqnarray}
%Eq (18) Zero modes in singular and regular gauges
we find the corresponding results for the correlators
\begin{eqnarray}
  \Pi (q)^{sing}
& = &
  \frac{N_{c}}{Q^{2}\,{\rho}^{4}\,2^{6}\pi ^{2}}
  \left[
          I_{1}(\xi )K_{0}(\xi )-I_{0}(\xi )K_{1}(\xi )
        + \frac{I_{1}(\xi )K_{1}(\xi )}{\xi}
  \right]^{2}_{\xi =\frac{Q\rho}{2}}
\nonumber \\
  \Pi (q)^{reg}
& = &
  \frac{N_{c}}{Q^{4}\,\rho ^{6}\,2^{6}\pi ^{2}}\,
  {\rm e}^{-2Q\rho} \; .
\label{E19}
\end{eqnarray}
%Eq (19) Correlators in terms of zero modes

Using now the integral representations
\begin{equation}
  K_{\nu}(\xi )
=
  \frac{(\frac{\xi}{2})^{\nu}\,
  \Gamma (1/2)}{\Gamma ({\nu} + 1/2)}
  \int_{1}^{\infty} {\rm e}^{-\xi t}\,(t^{2}-1)^{\nu - 1/2} dt \; ,
\label{E20}
\end{equation}
%Eq (20) Integral representation for K_\nu (\xi )
\begin{equation}
  I_{\nu}(\xi )
=
  \frac{(\frac{\xi}{2})^{\nu}}{\Gamma ({\nu} + 1/2)\Gamma (1/2)}
  \int_{0}^{\pi} {\rm e}^{\pm \xi\cos\theta}\,
  \sin ^{2\nu}(\theta ) \;d{\theta} \; ,
\label{E21}
\end{equation}
%Eq (21) Integral representation for I_\nu (\xi )
and the following Borel transforms, in accordance with Eq.~(\ref{E8}),
\begin{eqnarray}
  B[{\rm e}^{-a\sqrt{s}}]
& = &
  \frac{a}{\sqrt{4\pi}\tau ^{3}}\, {\rm e}^{-\frac{a^{2}}{4\tau ^2}}
\nonumber \\
  B\left[{\frac{1}{s^{2}}}\, {\rm e}^{-a\sqrt{s}}\right]
& = &
   \left( \tau ^{2} + \frac{a^{2}}{2} \right)
   \left[ 1-{\rm erf}\left(\frac{a}{2\tau}\right) \right]
 - \frac{{\tau}a}{\sqrt\pi}\, {\rm e}^{-\frac{a^{2}}{4\tau ^{2}}} \; ,
\label{E22}
\end{eqnarray}
%Eq (22) Borel transforms
it follows
\begin{equation}
  \Pi ^{sing}(\tau )
=
  \frac{N_{c}}{8(2\pi )^{4}\sqrt{\pi}\zeta\tau ^{4}}\,I \; ,
\label{E23}
\end{equation}
%Eq (23) Correlator in singular gauge: final result
where
\begin{equation}
  I
=
  \int_{0}^{\pi}d\theta_{1} \int_{0}^{\pi}d\theta_{2}
  \int_{1}^{\infty}dt_{1} \int_{1}^{\infty}dt_{2}\,
  C\,t\,{\rm e}^{-\frac{\zeta ^{2}\,t^{2}}{16}};
\label{24}
\end{equation}
%Eq (24) Integral I over t and $\theta$
with
$
 t
=
 \cos\theta _{1} + \cos\theta _{2} + t_{1} + t_{2}
$
and
\begin{eqnarray}
  C
=
& \phantom{} & \!\!\!\!\!
  \left(
           \frac{1}{8}
        -  \frac{1}{4}\sin ^{2}\theta _{1}
        + 2\sin ^{2}\theta _{1}\sin ^{2}\theta _{2}
  \right)
        \sqrt{t_{1}^{2}-1}\sqrt{t_{2}^{2}-1}
\nonumber  \\
& - &
        \frac{1}{4}\sin ^{2}\theta _{1}
        \cos ^{2}\theta _{2}\,
        \frac{\sqrt{t_{1}^{2}-1}}{\sqrt{t_{2}^{2}-1}}
      + \frac{1}{8}\sin ^{2}\theta _{1}\sin ^{2}\theta _{2}\,
        \frac{1}{\sqrt{t_{1}^{2}-1}\sqrt{t_{1}^{2}-1}} \; .
\end{eqnarray}
%Eq (25) Expression for C entering I
The values of the integral $I$ are tabulated below.
$$
\begin{tabular}{|l|l|l|l|l|l|}
\hline
$\zeta$ &   1 &  1.5 & 2   & 2.5 & 3    \\
\hline
$I    $ & 279 & 26.4 & 4.4 & 1   & 0.29 \\
\hline
\end{tabular}
$$

The analogous result to Eq.~(\ref{E23}) in regular gauge is
\begin{equation}
  \Pi ^{reg}(\tau )
=
  \frac{N_{c}}{16(2\pi )^{2}\zeta ^{6}\tau ^{4}}\,f(\zeta )
\label{E26}
\end{equation}
%Eq (26) Correlator in regular gauge: final result
with
\begin{equation}
  f(\zeta )
=
  \left( 1 + 2\zeta ^{2} \right)
  [1-{\rm erf}(\zeta )]
 -\frac{2\zeta}{\sqrt{\pi}}\,{\rm e}^{-\zeta ^{2}} \; ,
\label{E27}
\end{equation}
%Eq (27) Function $f(\zeta )$ entering $\Pi ^{reg}(\tau )$
where
\begin{equation}
  {\rm erf}(\zeta )
=
  \frac{2}{\sqrt{\pi}}\int_{0}^{\zeta}{\rm e}^{-y^{2}}dy \; .
\label{E28}
\end{equation}
%Eq (28) Definition of the function erf$(\zeta )$

Comparing these results with the effective single-instanton
contribution, given by Eq.~(\ref{E11}), at $\tau =\rho$, we deduce
\begin{equation}
  \Pi ^{sing}_{mult.}(\tau )
\approx
  \frac{7}{11}\,\Pi ^{sing}_{eff.}(\tau ) \; .
\label{29}
\end{equation}
%Eq (29) Equivalence between multi-instanton and effective
%        single-instanton contribution
Taking into account that the accuracy of the results obtained by
QCD sum rules is limited by uncertainties on the order of $30\%$,
we may claim that within the context of assuming the validity of
Eq.~(\ref{E4}), both approaches give coincident results in singular
gauges for the pion correlator $\Pi (\tau )$ at large $\tau$.
This means that in the region of large distances, multi-instanton
contributions to the correlator, obtained via summation over
planar diagrams~\cite{DP86,DP85} on one hand, and those from the
analysis~\cite{Shu83,SVZ80}, based on an effective single-instanton
approach on the other hand, are actually two different languages
which correctly describe the same phenomena. Perhaps even more
importantly, both methods achieve saturation of the QCD sum rules
only by including in the pseudoscalar channels the instanton
contribution. This conforms with the assumption that the QCD vacuum
is dominated by small-size instantons.

As regards the regular gauge, evaluation of $\Pi ^{reg}(\tau )$
(cf. (\ref{E26})) amounts to a very small value relative to
$\Pi ^{sing}(\tau )$ in the region where the QCD sum rules apply.
A strong cancellation in regular gauge at large distances was also
pointed out in~\cite{Hut95b}.
The investigation presented in this work provides further arguments
in favor of a singular gauge when processes at low energies (i.e.,
large distances) are studied.

\acknowledgments
The work of S.\ V.\ E. was supported in part by Grant 93-283.
One of us (N.\ G.\ S.) thanks the members of the Joint Institute for
Nuclear Research for the warm hospitality extended to him during his
stay.
\newpage  %finishes text part

\newpage  %finishes references

\begin{thebibliography}{99}
\bibitem{SVZ79} M.\ A.\ Shifman, A.\ I.\ Vainshtein, and
                V.\ I.\ Zakharov,
                Nucl.\ Phys.\ {\bf B147},\ 385\ (1979);
                {\bf B147},\ 448\ (1979);
                {\bf B147},\ 519\ (1979).
%[1]
\bibitem{Iof81} B.\ L.\ Ioffe,
                Nucl.\ Phys.\ {\bf B188},\ 317\ (1981);
                {\bf B191},\ 591(E)\ (1981).
%[2]
\bibitem{BI82} V.\ M.\ Belyaev and B.\ L.\ Ioffe,
               Zh.\ Eksp.\ Teor.\ Fiz.\ {\bf 83},\ 876\ (1982)
               [Sov.\ Phys.\ JETP\ {\bf 56},\ 493\ (1982)].
%[3]
\bibitem{CDKS82} Y.\ Chung\ {\it et al.},
                 Nucl.\ Phys.\ {\bf B197},\ 55\ (1982);
                 Z.\ Phys.\ C~{\bf 15},\ 367\ (1982).
%[4]
\bibitem{IS83} B.\ L.\ Ioffe and A.\ V.\ Smilga,
               Nucl.\ Phys.\ {\bf B216},\ 373\ (1983).
%[5]
\bibitem{NR82} V.\ A.\ Nesterenko and A.\ V.\ Radyushkin,
               Phys.\ Lett.\ {\bf 115B},\ 410\ (1982).
%[6]
\bibitem{NSVZ82}V.\ A.\ Novikov\ {\it et al.},
                Sov.\ J.\ Part.\ Nucl.\ Phys.\ {\bf 13},\ 542\
                (1982).
%[7]
\bibitem{Shu82} E.\ V.\ Shuryak,
                Nucl.\ Phys.\ {\bf B203},\ 93\ (1982);
                {\bf B203},\ 116\ (1982);
                {\bf B203},\ 140\ (1982).
%[8]
\bibitem{Shu83} E.\ V.\ Shuryak,
                Nucl.\ Phys.\ {\bf B214},\ 237\ (1983);
                Phys.\ Rep.\ {\bf 115},\ 151\ (1984).
%[9]
\bibitem{DP84} D.\ I.\ Diakonov and V.\ Yu.\ Petrov,
               Nucl.\ Phys.\ {\bf B245},\ 259\ (1984).
%[10]
\bibitem{DP86} D.\ I.\ Diakonov and V.\ Yu.\ Petrov,
               Nucl.\ Phys.\ {\bf B272},\ 457\ (1986).
%[11]
\bibitem{DP85} D.\ I.\ Diakonov and V.\ Yu.\ Petrov,
               Zh.\ Eksp.\ Teor.\ Fiz.\ {\bf 89},\ 361\ (1985);
               {\bf 89},\ 751\ (1985).
%[12]
\bibitem{CH92} M.\ C.\ Chu and S.\ Huang,
               Phys.\ Rev.\ D~{\bf 45},\ 2446\ (1992).
%[13]
\bibitem{ET89} S.\ V.\ Esaibegyan and C.\ N.\ Tamaryan,
               Yad.\ Fiz.\ {\bf 49},\ 815\ (1989);
               {\bf 55},\ 2193\ (1992);
               {\bf 58},\ 1\ (1995).
%[14]
\bibitem{ET95} S.\ V.\ Esaibegyan and C.\ N.\ Tamaryan,
               Pis'ma\ v\ JETF\ {\bf 61},\ 3\ (1995).
%[15]
\bibitem{Koc85} N.\ I.\ Kochelev,
                Yad.\ Fiz.\ {\bf 41},\ 456\ (1985)
                [Sov.\ J.\ Nucl.\ Phys.\ {\bf 41},\ 291\ (1985)];
                A.\ E.\ Dorokhov and N.\ I.\ Kochelev,
                Yad.\ Fiz.\ {\bf 52},\ 214\ (1990)
                [Sov.\ J.\ Nucl.\ Phys.\ {\bf 52},\ 135\ (1990)];
                A.\ E.\ Dorokhov, N.\ I.\ Kochelev, and Yu.\ Zubov,
                Sov.\ J.\ Part.\ Nucl.\ Phys.\ {\bf 23},\ 522\
                (1992);
                A.\ E.\ Dorokhov,
                Nucl.\ Phys.\ {\bf A581},\ 654\ (1995).
%[16]
\bibitem{DKZ93} A.\ E.\ Dorokhov, N.\ I.\ Kochelev, and Yu.\ Zubov.
                Int.\ J.\ Mod.\ Phys.\ {\bf A5},\ 603\ (1993);
                and references cited therein.
%[17]
\bibitem{DK90} A.\ E.\ Dorokhov and N.\ I.\ Kochelev,
               Z.\ Phys.\ C~{\bf 46},\ 281\ (1990).
%[18]
\bibitem{FB93} H.\ Forkel and M.\ K.\ Banerjee,
               Phys.\ Rev.\ Lett.\ {\bf 71},\ 484\ (1993).
%[19]
\bibitem{SVZ80} M.\ A.\ Shifman, A.\ I.\ Vainshtein, and
                V.\ I.\ Zakharov,
                Nucl.\ Phys.\ {\bf B163},\ 45\ (1980).
%[20]
\bibitem{DEK96} A.\ E.\ Dorokhov, S.\ V.\ Esaibegyan,
                and N.\ I.\ Kochelev,
                submitted to Yad.\ Fiz..
%[21]
\bibitem{Shu93} E.\ V.\ Shuryak,
                Rev.\ Mod.\ Phys.\ {\bf 61},\ 1\ (1993).
%[22]
\bibitem{Hut95a} M.\ Hutter,
%                ``Instantons in QCD and meson correlation
%                  functions''
                 M\"unchen preprint\ \#~LMU-95-01\ (1995).
%[23]
\bibitem{Hut95b} M.\ Hutter,
%                ``Gauge Invariant Quark Propagator in the Instanton
%                  Background''
                 M\"unchen preprint\ \#~LMU-95-03\ (1995).
%[24]
\end{thebibliography}
\end{document}